\documentclass[9pt,twocolumn,twoside]{opticajnl}
\journal{opticajournal} 

\setboolean{shortarticle}{true}

\usepackage{lineno}
\linenumbers 

\title{Metal-Mesh Linear Variable Bandpass Filter for Far-Infrared Wavelengths}

\author[1*]{Joanna Perido}
\author[2]{Kevin Denis}
\author[3]{Sean O. Clancy}
\author[2]{Nicholas F. Cothard}
\author[4]{Peter K. Day}
\author[2]{Jason Glenn}
\author[4]{Henry Leduc}
\author[2]{Manuel Quijada}
\author[2]{Jessica Patel}
\author[2]{Edward Wollack}

\affil[1]{University of Colorado Boulder}
\affil[2]{NASA Goddard Space Flight Center}
\affil[3]{HZO, Inc., 5151 McCrimmon Parkway, Suite 208, Morrisville, NC 27560} 
\affil[4]{Jet Propulsion Laboratory}
\affil[*]{joanna.perido@colorado.edu}

\begin{abstract}
Future far-infrared (IR) observatories require compact and cost efficient optical linear variable bandpass filters (LVBFs) to define their instrument spectral bands. We have designed novel far-IR LVBFs that consist of metal-mesh bandpass filters comprised of a gold film with cross-slots of varying sizes along a silicon (Si) substrate with anti-reflection (AR) coatings. We present our work on the simulated and measured transmission of non-AR coated and AR coated LVBFs for bandpass peaks from wavelengths of 24 to 36 $\mu$m with a resolving power ($R=\lambda_0/\Delta\lambda$) of R$\approx$6 for non-AR coated LVBFs and R$\approx$4 for AR coated LVBFs. We also present a method to decrease the effects of out-of-band high frequency transmission exhibited by metal-mesh filters by depositing a thin layer of hydrogenated amorphous silicon (a-Si:H) on the metal-mesh of the LVBF. We have fabricated and measured the LVBFs at room temperature and cryogenic temperatures (5 K). We measure a high peak transmission of $\sim$80-90 \% for the AR coated LVBF at 5 K and demonstrate that the a-Si:H LVBF is a promising method to address out-of-band high frequency transmission. 
\end{abstract}

\setboolean{displaycopyright}{false} 

\begin{document}
\nolinenumbers

\maketitle

\section{Introduction}
\begin{figure}[h!]
\centering
\includegraphics[width=\linewidth]{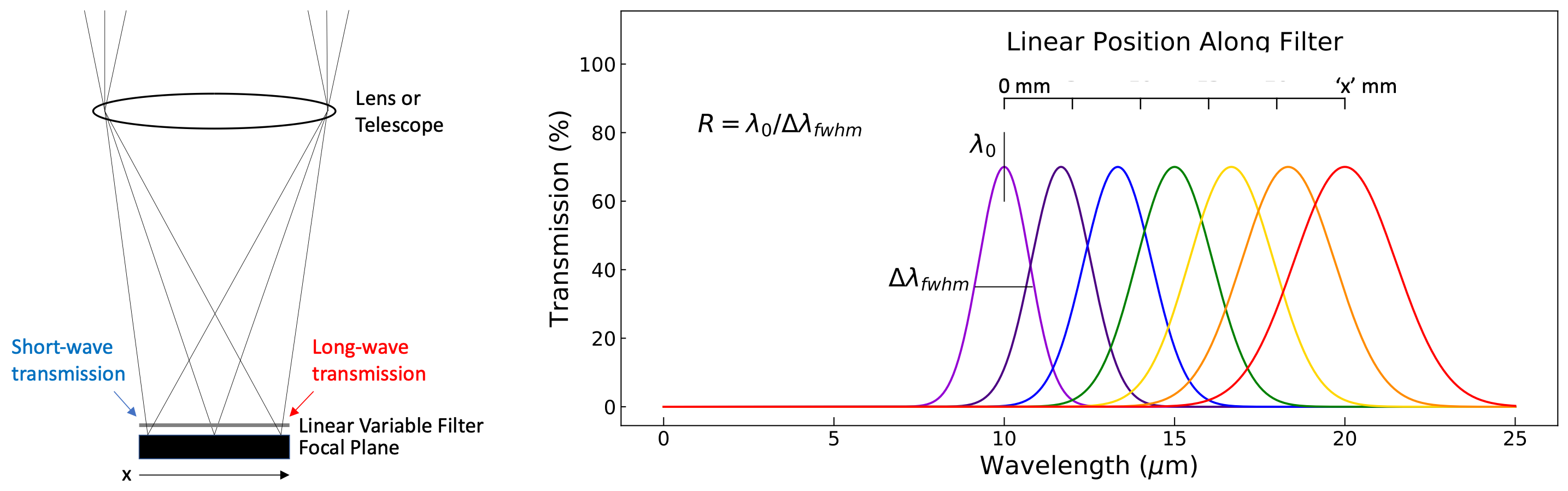}
\caption{\textit{Left}: A schematic showing how a continuous LVBF is placed in an imaging optical system to create a spectral mapper. The LVBF is placed directly in front of the focal plane array. \textit{Right}: A schematic of the spectral transmission for the LVBF. The bandpass central wavelength $\lambda_0$ varies linearly along the filter length and the resolving power ($R$) stays constant.}
\label{fig:LVF_Cartoon1}
\end{figure}

The proposed far- infrared (IR) sub-orbital observatory the Balloon Experiment for Galactic INfrared Science (BEGINS) will utilize a cryogenic instrument to map the spectral energy
distributions (SEDs) of interstellar dust in the vicinity of high-mass stars to measure electromagnetic radiation fields and dust properties in a variety of environments \cite{perido2023kinetic}.
These observations require optical filters to define the instrument bands with a specific resolving power (or spectral resolution) $R=\lambda_0/\Delta\lambda$, where $\lambda_0$ is the bandpass peak wavelength. Our goal is to define the instrument bands of BEGINS with linear variable bandpass filters (LVBFs) at the focal plane of the instruments (Fig. \ref{fig:LVF_Cartoon1}). LVBFs are filters with bandpasses that vary linearly along their length. They are crucial for many applications that range from astronomy to pharmaceutical analysis to imaging sensors for piloted aircrafts, to name a few \cite{emadi2012linear,grossman2007spectral, serruys2014linear, williams2016single}. The LVBFs on BEGINS will enable hyperspectral imaging from 25 to 65 $\mu$m with a lower limit resolving power of R=7.5. This resolving power is significant, because at R$\geq$7.5 the effects of dust grain size and radiation field intensity from 25 to 65 $\mu$m can be separated \cite{perido2023kinetic}. This will allow astronomers to confirm theoretical work on the predicted shapes of SEDs within this spectral region. 


\par
\begin{figure}[h!]
\centering
\includegraphics[width=\linewidth]{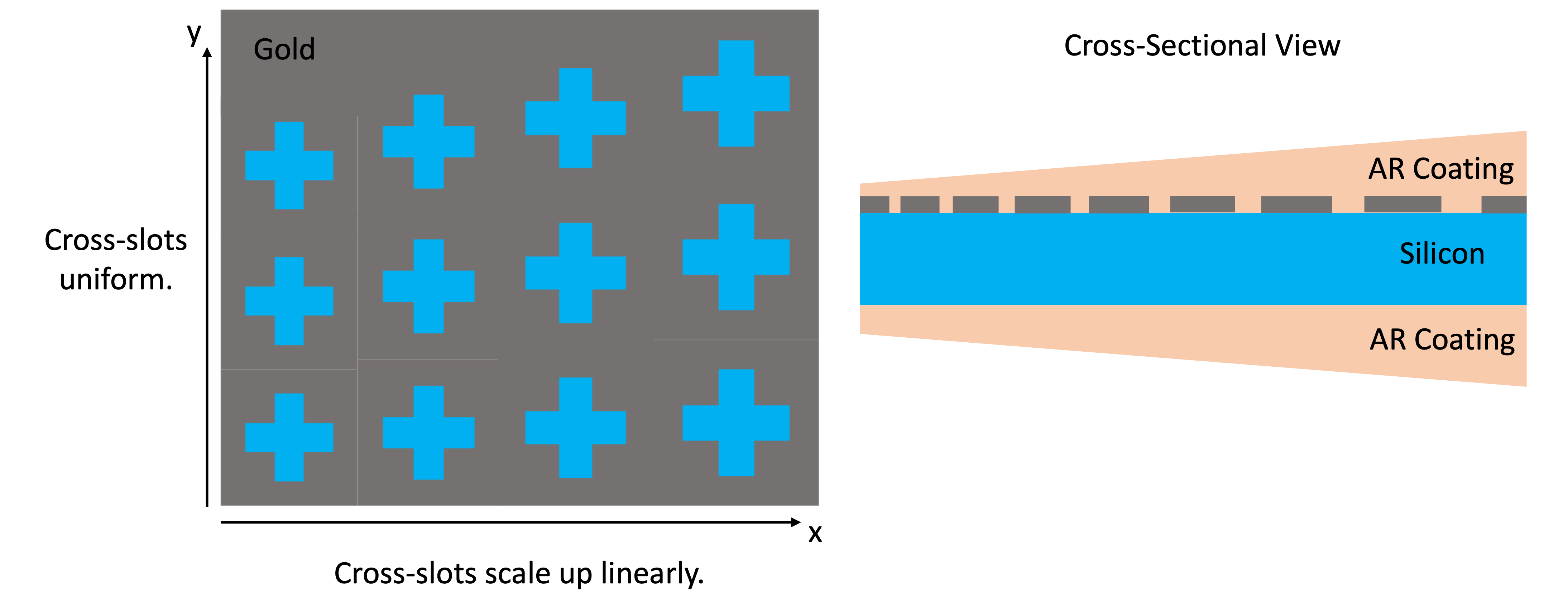}
\caption{Figures not to scale. \textit{Left:} A schematic showing how the cross-slot geometry scales linearly with increasing bandpass wavelength along the filter. The gray represents the gold film and the blue represents the bare Si. \textit{Right:} A cross-sectional view of the filter. The AR-coatings linearly increase in thickness by $\lambda_0/4\sqrt{\epsilon_{AR}}$ along the length of the filter.}
\label{fig:LVF_Cartoon2}
\end{figure}

 The LVBFs will be comprised of metal-mesh bandpass filters. Metal-mesh filters have been studied for far-IR instruments since the first publication by Ulrich \cite{ulrich1967far}. They were chosen for their simple fabrication scheme, cost efficiency, and compactness. The simplest form consists of a single layer of metal-mesh that can be free-standing or deposited on a substrate. The mesh consists of a periodic structure, the geometry of which determines whether the filter is a low-pass, high-pass or bandpass filter. The LVBFs for BEGINS will be comprised of thin film gold with cross-slot dimensions scaled linearly in all dimensions along the length of a silicon (Si) substrate (Fig. \ref{fig:LVF_Cartoon2}). A cross-slot geometry has both inductive and capacitive properties that make it self-resonant, leading to a bandpass response \cite{porterfield1994resonant, merrell2012compact, ade2006review}. Its response can be easily tailored by changing the cross-slot dimensions, which include its cross-length, cross-width and periodicity (or the distance from the center of one cross to the other, also referred to as the cross-pitch) \cite{porterfield1994resonant}. The bandpass peak wavelength increases along the length of the LVBF, while preserving the resolving power, by linearly scaling the cross-slot dimensions. The cross-slot dimensions and their effect on the shape of the bandpass and its peak wavelength are discussed in further detail in Sec. \ref{sec:filt_design}. Gold was chosen because it has a low resistivity, which minimizes ohmic losses and allows for high transmission. We use high-resistivity floatzone Si wafers for low dielectric loss. To reduce the reflective losses at the vacuum-Si interfaces, Parylene-C anti-reflection (AR) coatings are deposited on both sides of the filter. The AR coatings linearly increase in thickness by $\lambda_0/4\sqrt{\epsilon_{AR}}$ along the length of the filter, where $\epsilon_{AR}$ is the relative permittivity of the AR coating. Parylene-C is a thermoplastic polymer that is commonly used in the far-IR spectral region as an AR coating due to its good adhesion, mechanical stability, minimal molecular out-gassing, and low water absorption. It has also proven to be cryogenically robust, withstanding repeated cooling cycles \cite{cothard2024monolithic}. We present the results of a non-AR coated and an AR-coated LVBF.

Although there are many advantages to metal-mesh filters, they exhibit out-of-band transmission at frequencies greater (or wavelengths shorter) than the bandpass peak, which we refer to as higher-order side bands. They occur because the periodic cross-slots act as a diffraction grating that diffracts a wave when the lattice size (or periodicity) is electrically large. Therefore, the frequency at which these higher-order side bands appear depends on the size of the cross-pitch. A common method used to suppress the high frequency higher-order side bands are stacked filters with polyimide in between the layers \cite{ade2006review, smith2003designer}. But this leads to an overall decrease in the bandpass throughput due to losses in the filter stack. It is also possible to filter them out using a low-pass filter. However, with a cross-slot metal-mesh LVBF the transmission of the higher-order side bands of the lower-frequency (longer-wavelength) bandpasses overlap with the higher-frequency (shorter-wavelength) bandpasses along the LVBF. In other words, the transmission of higher-order side bands and bandpasses on the LVBF will occur at the same frequencies. Therefore, it is not possible to filter out the higher-order side bands without also filtering out the bandpasses of the LVBF. To address this issue we investigate a method to increase the spectral distance between the bandpass peak and first higher-order side band. We do this by depositing a thin layer of hydrogenated amorphous silicon (a-Si:H) on the metal-mesh layer. The addition of the a-Si:H with a high permittivity compared to free space allows for the cross-pitch to be decreased, resulting in higher-order side bands that are shifted further away from the bandpass peak to higher frequencies. This occurs because the capacitance of the metal-mesh changes when it is immersed in the a-Si:H. In order to achieve the same bandpass peak as before the addition of the a-Si:H the cross-slot dimensions must be decreased. If the higher-order side bands of each bandpass peak are shifted far enough we can use a low-pass filter and avoid filtering out the bandpasses.
%

In this paper, we present our work on the development and measurement results of a prototype non-AR coated and AR coated LVBF and a a-Si:H LVBF with targeted bandpasses that vary from 24 to 36 $\mu$m. In Sec. \ref{sec:filt_design} we discuss the filter design and modeling scheme to simulate the transmission of the filter. In Sec. \ref{sec:fabrication} we describe the fabrication method and deviations from design of the fabricated LVBF. In Sec. \ref{sec:MeasDesc} we explain the filter transmission measurement method. In Sec. \ref{sec:LVFresults} we discuss the LVBF transmission measurements and compare the measurements to simulations. In Sec. \ref{sec:asi_LVF} we discuss the results of our initial investigation on a a-Si:H LVBF. In the last section, Sec. \ref{sec:Con}, we add concluding remarks and discuss future work on the LVBFs. 

\section{Filter Design and Filter Modeling}
\label{sec:filt_design}
\par
\begin{figure}[h!]
\includegraphics[width=\linewidth]{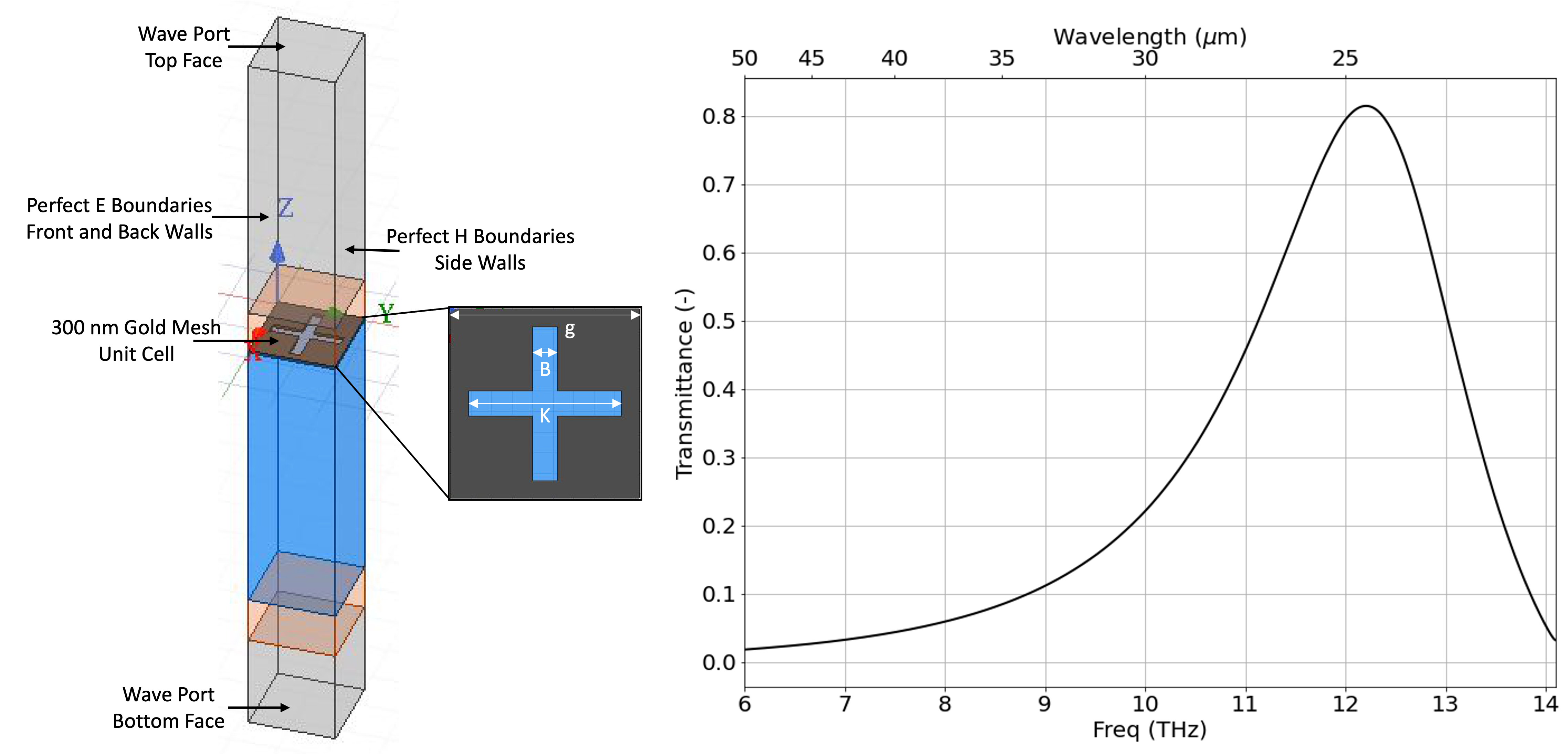}
\caption{\textit{Left}: 3D model of an HFSS unit cell simulation for an AR coated metal-mesh bandpass filter. For illustration purposes, the vacuum, Si substrate and AR coating box heights are not to scale. The cross-slot dimensions are defined by the periodicity (g), the cross-length (K), and the cross-width (B). \textit{Right}: The calculated transmittance spectrum of a metal-mesh bandpass filter from HFSS simulation results. The peak transmittance, resolving power, and bandpass peak wavelength are set by the cross-slot dimensions.}
\label{fig:Sim_Example}
\end{figure}

 The transmission profile of a metal-mesh bandpass filter is similar to a Lorentzian and is determined by its cross-slot dimensions: the cross-pitch (g), the cross-length (K) and the cross-width (B) (Fig. \ref{fig:Sim_Example}) \cite{tomaselli1981far, porterfield1994resonant, moller1996cross, melo2012cross, merrell2012compact}. The bandpass peak scales with K (the cross-length) and the bandwidth becomes small as the ratios of g/K and g/B increase \cite{porterfield1994resonant}. In order to predict and model the metal-mesh bandpass filter performance we used Ansys High Frequency Structure Simulator (HFSS) software \cite{ansys}. HFSS is a full-wave frequency domain electromagnetic field solver based on the finite element method. It numerically solves Maxwell's equations across a specified frequency range for a specified structure geometry, material configuration, and boundary conditions. HFSS is used to extract S-parameters and predict the transmission profiles of the metal-mesh filters. Through symmetry, an array of uniform cross-slots in a gold film on a Si substrate can be simulated by a single unit cell with perfect electric ($\vec{E}$) and magnetic ($\vec{H}$) field boundary conditions \cite{merrell2012compact, porterfield1994resonant}. The unit cell structure is shown in the left panel of Fig. \ref{fig:Sim_Example}. Wave ports are used to excite and monitor the unit cell response for a normally incident wave. A quarter of the unit cell mesh geometry is simulated and layers representing vacuum, the metal-mesh, and the substrate are incorporated in filter stack model. A wave port is equivalent to a semi-infinite waveguide that supports the modes $TE_{00}$ and $TM_{00}$.

\par The cross-slot dimensions are initially calculated from the desired bandpass peak wavelength ($\lambda_0$). The cross-length, parameter K, is approximately $\lambda_0/2$, where $\lambda_0$ is the bandpass peak wavelength in the medium, which in this case is Si ($n_{Si}$ = 3.42)  \cite{melo2012cross,moller1996cross}. The estimated dimensions are then optimized using simulations to achieve the desired resolving power and maximum transmission. The metallization is modeled as a 300 nm thick layer with an effective bulk conductivity of $3.33\times10^7$ Siemens/m. The conductivity was determined from room temperature DC resistivity measurements made at Goddard Space Flight Center (GSFC) on previous filter samples fabricated with 300 nm gold for this project. After fabrication of the LVBFs, the actual room temperature conductivity of the filters are determined from DC resistivity measurements and used to properly determine the fabricated bandpass transmission profile. The cryogenic bulk conductivity is calculated using the measured DC residual resistivity ratio (RRR) of the metallization, which is the ratio of the electrical resistance of a metal at room temperature and 4.2 K. The Si in the model is defined as a 527 $\mu$m thick volume with a relative permittivity of $\epsilon_{Si}$ = 11.7, which was determined through room temperature measurements on a high-resistivity float zone Si wafer sample at GSFC. Through simulations the Si dielectric loss was determined to be negligible. There is $\sim$0.3 reflectance at each vacuum-Si interface, so we implemented quarter-wave AR coatings on both the metal-mesh side and on the back side of the filter to increase transmission. Since we use Parylene-C for the AR-coatings, the relative permittivity is set to $\epsilon_{AR}$ = 2.6 and a quarter-wave layer is adopted for the thickness (= 3.72 $\mu$m for $\lambda_0=24$ $\mu$m), which provides a reasonable match between the silicon and vacuum \cite{gatesman2000anti}. Simulations without the AR-coatings were performed to determine the cross-slot dimensions for a bandpass peak at 24 $\mu$m. The dimensions were then scaled up to 36 $\mu$m to span the full bandpass wavelength range of a LVBF that is 17 mm in length. The simulation optimized cross-slot dimensions for the bandpass peak at 24 $\mu$m and scaled up cross-slot dimensions for the bandpass peak at 36 $\mu$m are listed in Table \ref{tab:cross_params}.

\begin{table}[h!]
    \centering
    \begin{tabular}{|c|c|c|}
        \hline
         Parameter ($\mu$m) & $\lambda_0= 24$ $\mu$m & $\lambda_0= 36$ $\mu$m\\
         \hline
         \hline
         g & 6.17 & 9.25\\
         \hline
         K & 4.95 & 7.43\\
         \hline
         B & 0.80 & 1.20\\
         \hline
    \end{tabular}
    \caption{The cross-slot dimensions for bandpass peaks at 24 $\mu$m and 36 $\mu$m for a 17 mm long LVBF. The cross-slot dimensions for the bandpass peak at 24 $\mu$m were optimized through simulations and scaled up to 36 $\mu$m to span the full bandpass wavelength range of the LVBF.}
    \label{tab:cross_params}
\end{table}

\section{Fabrication}
\label{sec:fabrication}

The LVBF was fabricated in a simple single layer process. Double side polished (DSP) intrinsic float zone silicon wafers ($\rho$ $>$ 20~k$\Omega$-cm) were coated with a 5~nm thick titanium (Ti) adhesion layer and 300~nm thick gold layer by electron beam evaporation in the GSFC Detector Development Laboratory (DDL). The cross-slots had minimum features of 0.8 $\mu$m which were lithographically patterned by a Heidelburg DWL 66+ direct write laser system and a single layer of S1805 resist. The gold was etched by argon ion milling (4-Wave) and the Ti was further etched by a combination of fluorine plasma and hydrofluoric acid. Several filters were fabricated on a single 100 mm silicon wafer. After etching, the photoresist was removed by oxygen plasma and solvent cleaning. The filters were then diced into 1 inch samples. The sheet resistance of the gold was measured to be 2.94 $\mu \Omega$-cm at 300 K with a RRR $\sim$ 5. A few of the LVBF samples were sent to HZO for the Parylene-C AR coating deposition \cite{hzo}. They developed a method to deposit a coating with a gradient thickness across the filter. Their deposition method is described in the following subsection. This way the thickness varies to approximately $\lambda_0$/4 of the bandpass peak across the filter. 

\par
\begin{figure}[h!]
\includegraphics[width=\linewidth]{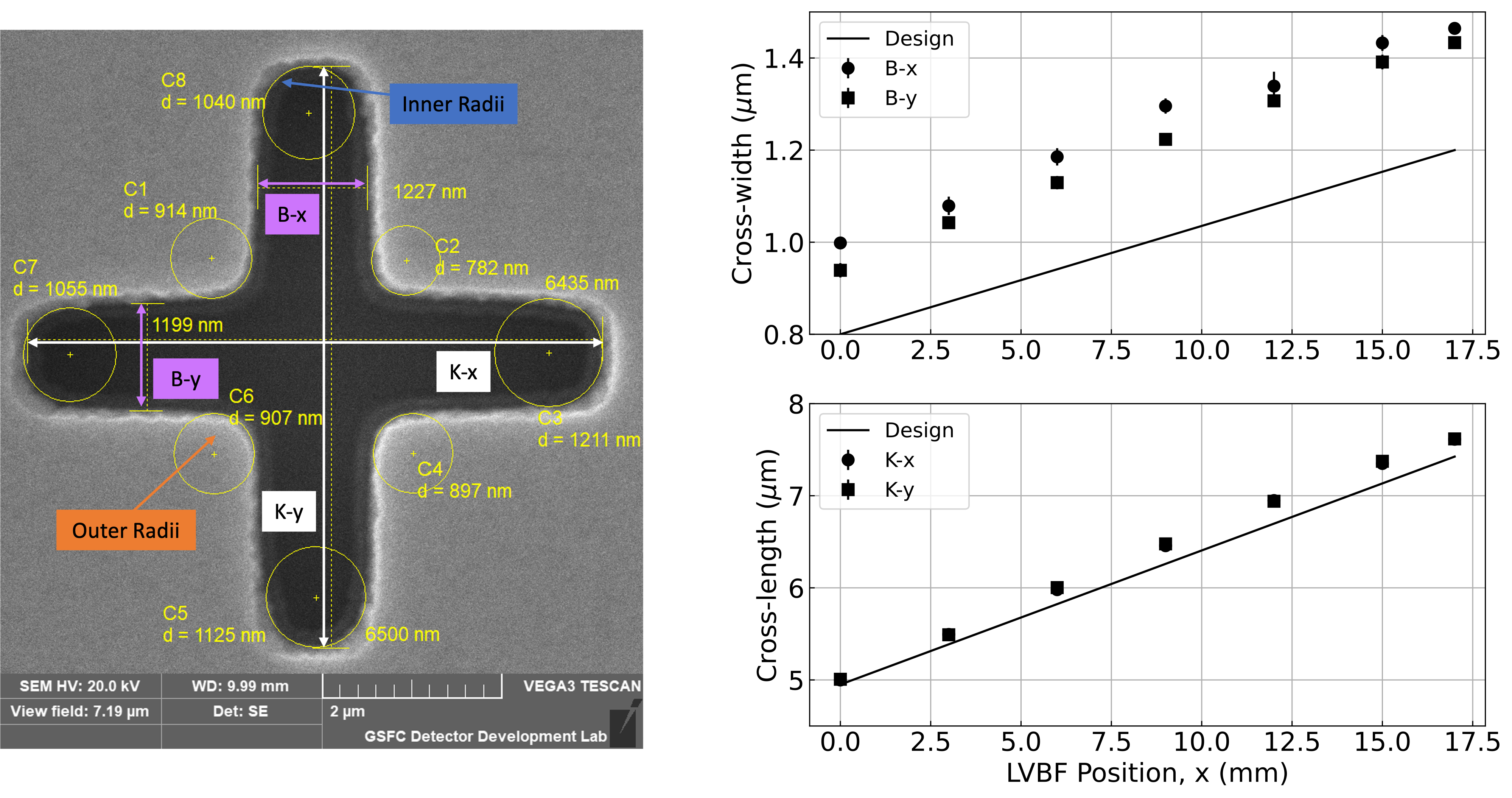}
\caption{\textit{Left}: SEM image of the cross-slot at (x, y) = (9, 8.5) mm. \textit{Right}: Plots showing the results of the fabricated cross-slot dimensions as a function of the LVBF position, where x = 0 mm corresponds to the $\lambda_{short}$ bandpass peak at the beginning of the filter. The top plot shows the measurements of the cross-width in the horizontal (B-x) and vertical (B-y) directions. The bottom plot shows the measurements of the cross-length in the horizontal (K-x) and vertical (K-y) directions. The solid line in the plots are the targeted design dimensions. Measurements were made at y = 3, 8.5, and 12 mm, which belong to the same bandpass peak column at the specified x-position along the length of the filter. The black markers are the average of the measurements made at each location with error bars corresponding to the standard deviation of the mean for the three measurements.}
\label{fig:LVF_fbdResults}
\end{figure}

 Measurements of the cross-slots were made on an LVBF sample after fabrication. The measurements were made at y = 3, 8.5, and 12 mm and at positions x = 0, 3, 6, 9, 12, 15 and 17 mm, where the bandpass varies along the x-axis. The left image in Fig. \ref{fig:LVF_fbdResults} shows a scanning electron microscope (SEM) image of the cross-slot at (x,y) = (9, 8.5) mm with an example of how the measurements were made. The results of the measurements are shown in the right panel of Fig. \ref{fig:LVF_fbdResults}. The top plot shows that the cross-widths deviated significantly from the design cross-widths with average errors of 25\% and of 20\% in ``B-x" and ``B-y", respectively. The bottom plot shows the measured cross-lengths were similar in the horizontal (K-x) and vertical (K-y) directions and had an average error of 3\% when compared to the design cross-length. The SEM image in Fig. \ref{fig:LVF_fbdResults} shows rounding of the inner and outer cross corners, which was also seen in other cross-slots across the filter. Measurement of the inner and outer radii were made along one row of the filter at  y = 8.5 mm and x = 0, 3, 6, 9, 12, 15 and 17 mm . The measurement was repeated four times at each x-position and averaged. On average the inner radii were determined to be 470 nm $\pm$ 110 nm and the outer radii were 375 $\pm$ 54 nm.

\par Since the transmission profile is sensitive to changes in the cross-slot dimensions \cite{melo2012cross, page1994millimeter, perido2022cross}, modifications were made to the simulations to determine how the bandpass peak would shift. The modifications included simulating the transmission profile of the cross-slot shown in the SEM image in Fig. \ref{fig:LVF_fbdResults} with the measured gold resistivity of the filters ($R_s$=2.94 $\mu \Omega$-cm). Incorporating this perturbation shifted the simulated bandpass peak from 24 $\mu$m to 23.82 $\mu$m. Therefore, based upon the realized geometry the bandpass peak is expected to increase across the filter from 23.82 $\mu$m to 35.74 $\mu$m.

\subsection{Parylene-C AR Coating Deposition Description}
\label{sec:HZO_desc}
The Parylene-C AR coating was deposited onto both sides of the LVBF using a chemical vapor deposition polymerization (CVDP) process. In this process all exposed surfaces will have completely conformal coatings. To achieve the desired coating gradient thicknesses, a series of double-sided gradient coating (DSGC) fixtures were designed through an iterative rapid prototyping process that evaluated the restrictive volume features above a silicon die’s surface and the deposited Parylene-C coating thickness \cite{clancy2024fixtures}.
\par
\begin{figure}[h!]
\center
\includegraphics[width=.8\linewidth]{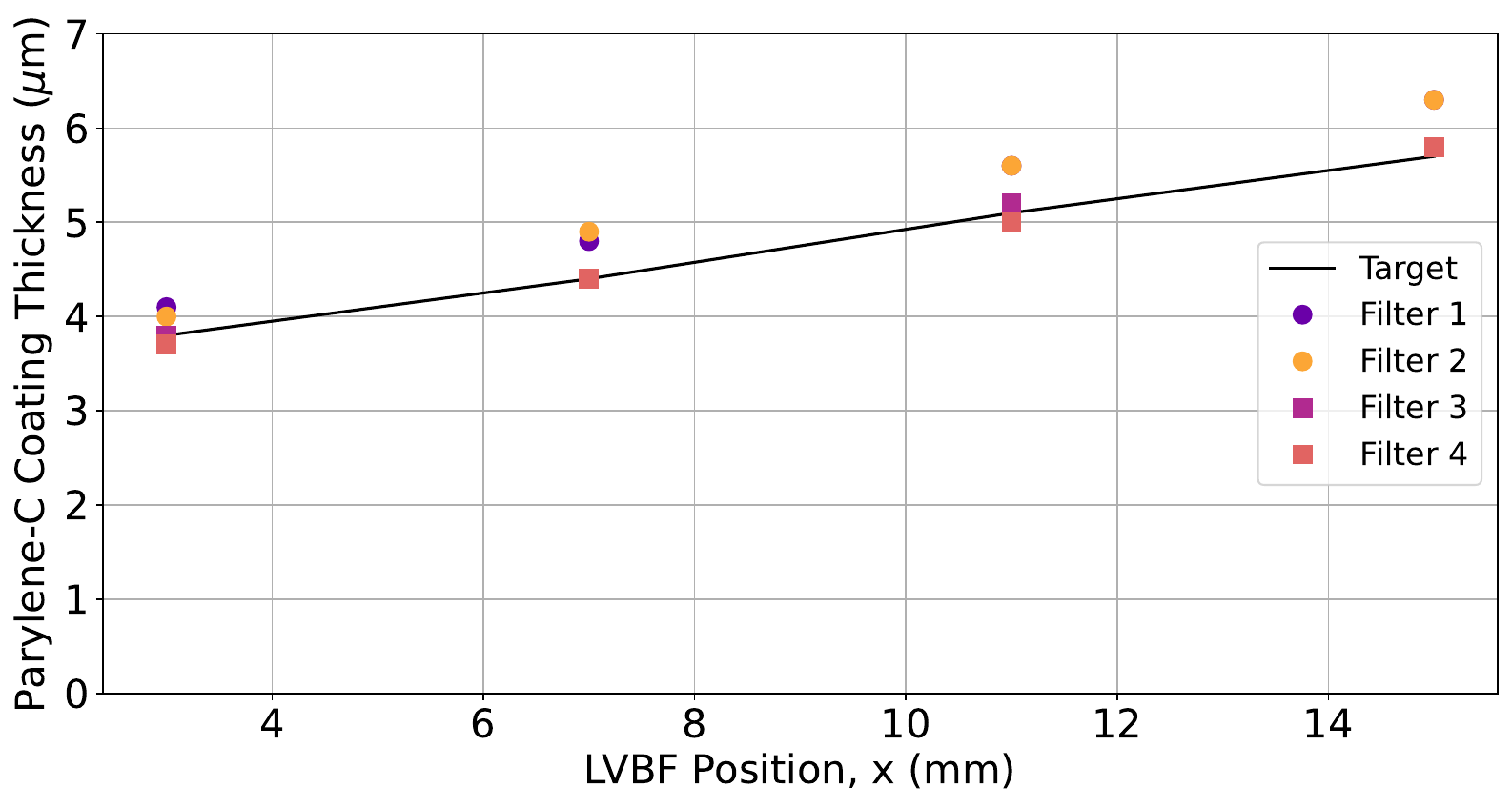}
\caption{Parylene-C AR coating thickness measurements at x = 3, 7, 11, and 15 mm. Black Line: Target thickness. The thickness linearly increases by $\lambda_0/4\sqrt{\epsilon_{AR}}$ along the length of the 17 mm LVBF. The coatings on both sides of the LVBFs were measured and averaged. Circles: measurements for filters without the Silane A-174 AP (Filter 1 and Filter 2). Squares: measurements for filters with the Silane A-174 AP (Filter 3 and Filter 4).}
\center
\label{fig:coat_thick}
\end{figure}

 The wafer-loaded fixtures, along with glass slide witness coupons, were placed in a Parylene deposition chamber. The Parylene-C precursor, commonly referred to as a dimer, was loaded into the deposition system’s vaporizer furnace. The dimer was sublimed under vacuum with a temperature ramp up to 160 °C to form a Parylene vapor, which was pulled by vacuum through a pyrolysis furnace heated to 620 °C, which cleaved the dimer into two reactive monomers. The monomers travel to the room temperature deposition chamber, where they coat every exposed surface within the chamber to form the Parylene-C polymer film with an overall growth rate of $\sim$1 µm/hour. Half of the final filters were coated with a Silane A-174 adhesion promoter (AP) and half without it. This was done in case the AP negatively affected the filter’s spectroscopic or imaging performance, though by its nature, the Silane A-174 AP significantly improved adhesion of the coating to the Si wafer. The final set of coating runs were split into two sets with the first set using Parylene-C precursor and no Silane A-174 AP and the second set using Parylene-C precursor and Silane A-174 AP. The thickness of the Parylene-C coating for these processes were defined by the amount of Parylene-C precursor loaded in the Parylene deposition system. The Parylene-C coating thickness was then verified with a Filmetrics F40 microscope-based spectral reflectance measurement tool \cite{film}. The coatings on both sides of the LVBFs were measured and averaged. Fig. \ref{fig:coat_thick} shows the results of the thickness measurements along the LVBF . The circles represent measurements for filters without the Silane A-174 AP (Filter 1 and Filter 2) and the squares represent the measurements for filters with the Silane A-174 AP (Filter 3 and Filter 4). The LVBF results discussed in Sec. \ref{sec:LVFresults} are from measurements made on Filter 4, which was close to the target coating thickness. 

\section{Filter Measurement Description}
\label{sec:MeasDesc}

\par The transmission measurements were made using a Bruker Optics – IFS 125HR, which is a high resolution Fourier transform infrared spectrometer (FTS). Measurements were made at 5 K and at room temperature in a focused beam with a 2 mm diameter illuminating aperture over a spectral range of 29 cm\textsuperscript{-1} to 648 cm\textsuperscript{-1}, with a spectral resolution of 2 cm\textsuperscript{-1}. The filters were placed in a sample holder. A 2 mm aperture was placed in front of the holder to control the beams size. The sample holder with both the filter and an open aperture was attached to a rod which moved through the optical path. First, a reference spectrum was collected for the open aperture (without filter). Then, for the linear variable filters the rod was moved down manually in segments of 3 mm to measure the transmission of the varying bandpasses along the filters. The transmission spectra was then calculated by taking the ratio of the beam spectrum going through the filter divided by the reference beam spectrum going through the hole.

\section{Filter Measurement vs Simulations: Results and Discussion}
\label{sec:LVFresults}

\begin{figure}[h!]
\includegraphics[width=\linewidth]{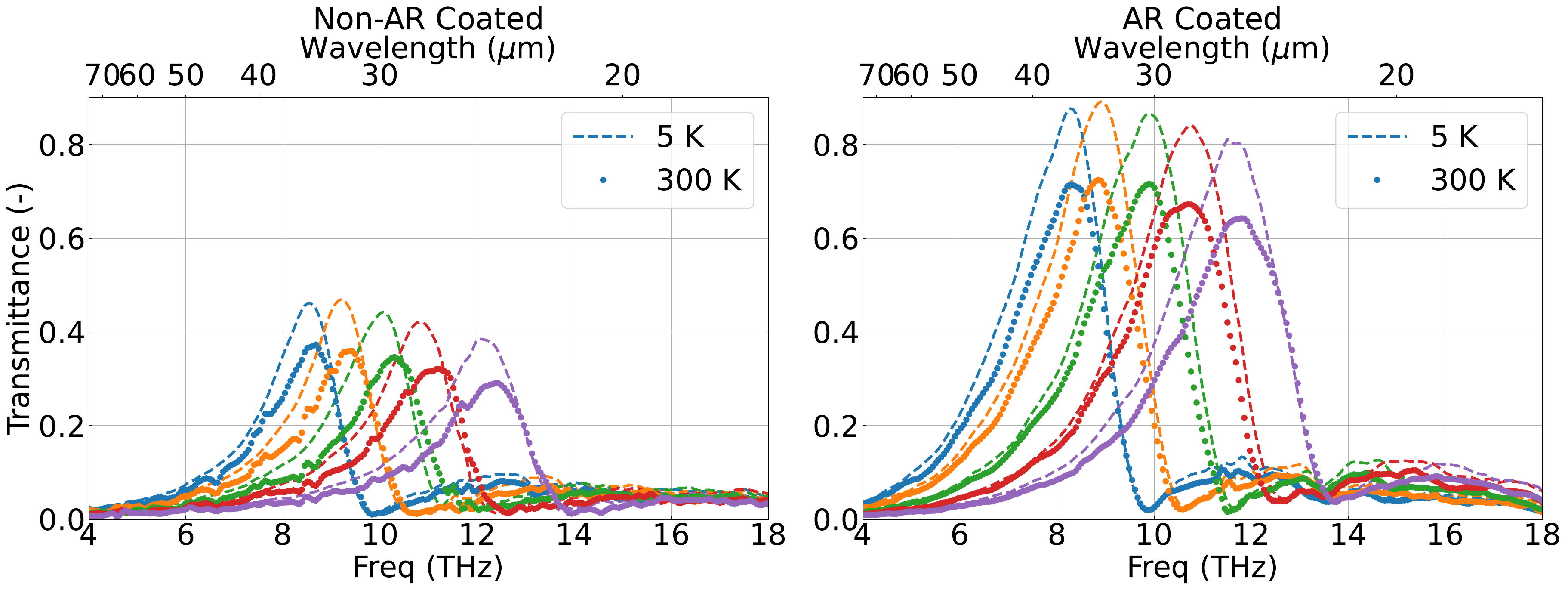}
\caption{FTS measured transmission spectra of LVBF at x = 3, 6, 9, 12 and 15 mm. \textit{Left}: Measurements of the non-AR coated LVBF at 300 K (dotted lines) and 5 K (dashed lines). \textit{Right}: Measurements of the AR coated LVBF at 300 K (dotted lines) and 5 K (dashed lines).}
\label{fig:LVF_meas}
\end{figure}

Room temperature (300 K) and cryogenic temperature (5 K) FTS measurements  were made of two different LVBF samples. One sample was non-AR coated and the other was AR coated on both sides with Parylene-C. Fig. \ref{fig:LVF_meas} shows the measured transmission across the LVBF at locations x = 3, 6, 9, 12, 15 mm with error bars of $\pm$1.25 mm. The error bars are derived from the finite aperture size and the uncertainty in the location of the aperture, which is 0.5 mm. In cooling from 300 K to 5 K, the non-AR (and AR) coated samples increase in peak transmission by $\sim$10\% ($\sim$17\%), whereas the model simulations, performed as described in Sec. \ref{sec:filt_design}, predict an increase of 7\% (10\%) for these sample configurations. The addition of AR coatings increases the peak transmission by 35\% at 300 K and 42\% at 5 K, whereas simulations predict an increase in peak transmission of 38\% at 300 K and 41\% at 5 K. The measurements are in agreement with simulations for the increase in transmission when the filters are cooled and AR-coated.

\par Another feature to note in the measurements are the higher-order side bands seen at frequencies higher than the bandpass peaks in Fig. \ref{fig:LVF_meas}. The location of the higher-order side band frequency cut-offs, where the transmission is nearly zero in the plot, can be predicted using Floquet's Theorem \cite{celozzi2023frequency}. For a 2D structure these solutions are referred to as Transverse Electric ($TE_{mn}$) or Transverse Magnetic ($TM_{mn}$) modes. For a normal incident plane wave the modes have frequency cut-offs that are determined by the following equation \cite{celozzi2023frequency},

\begin{equation}
    \label{floqetmode_cutoff}
    f_{c,mn} = \frac{c}{2\pi\sqrt{\epsilon_r}}\sqrt{\bigg(\frac{2\pi m}{g_x}\bigg)^2+\bigg(\frac{2\pi n}{g_y}\bigg)^2},
\end{equation}

\noindent where $c$ is the speed of light, $\epsilon_r$ is the relative permittivity of the medium, which in our case is Si, and $g_x$ and $g_y$ are the cross-pitch in the x and y direction. This equation is only valid and used for the non-AR coated filters that we compared to experimental measurements. A modified equation would be required for the AR-coated and a-Si:H LVBFs, since their relative permittivity also needs to be considered. Since the transmission was only measured up to 19 THz only one higher-order side band is shown for each bandpass after the frequency cutoff for Floquet modes $TE_{10/01}$ and $TM_{10/01}$. The frequency cutoffs for the non-AR coated sample are expected to be 9.9 THz, 10.6 THz, 11.7 THz, 12.7 THz, and 14.0 THz for each measured bandpass peak. There is excellent agreement between predicted and measured cutoff frequencies, which suggests that the fabricated cross-pitch is consistent with the design cross-pitch. Fig. \ref{fig:LVF_meas} \textit{Right} shows that the addition of the AR coating causes the higher-order side bands to increase in transmission but fractionally less than the increase in transmission of the bandpass peak, where the quarter-wave AR coating is more effective. As mentioned in the introduction, the out-of-band transmission is undesirable because the transmission of the higher-order side bands of the lower-frequency bandpasses overlaps with the higher-frequency bandpasses. We discuss a first attempt to address this problem in Sec. \ref{sec:asi_LVF}.

\par
\begin{figure}[h!]
\includegraphics[width=\linewidth]{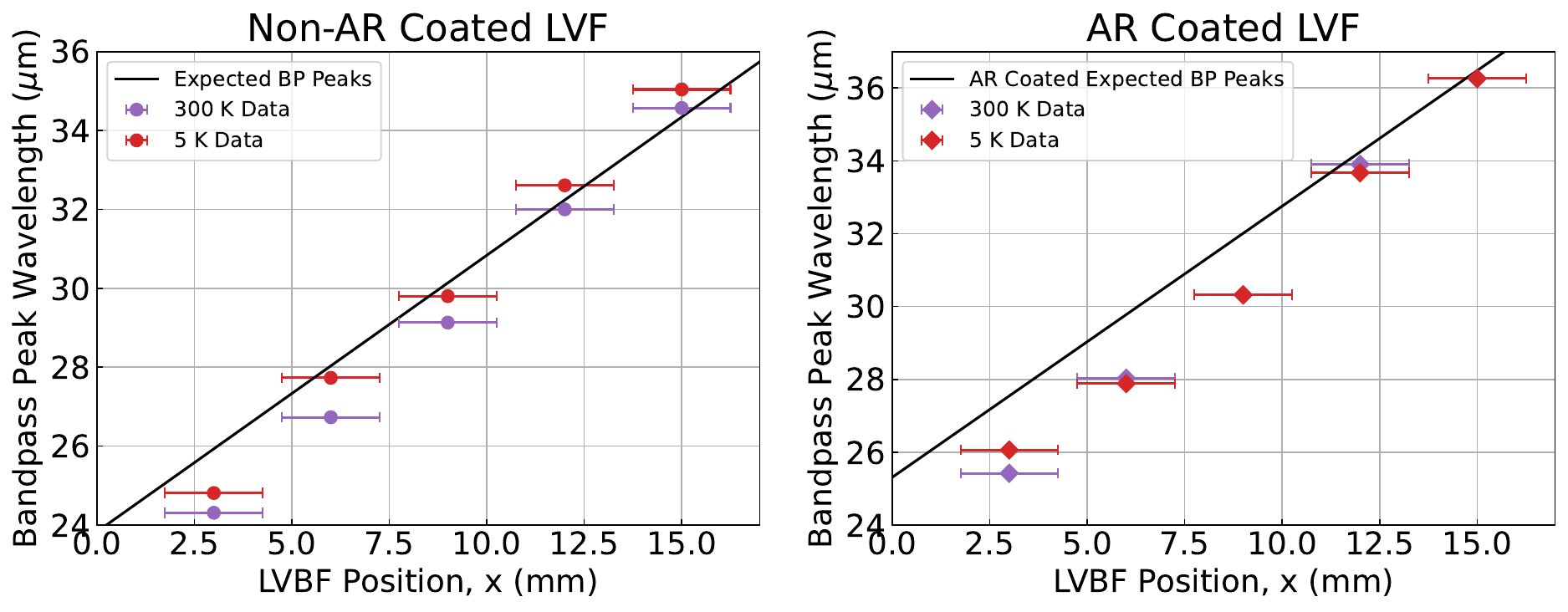}
\caption{Plots of the wavelength of the bandpass peak as a function of the LVBF position, where x = 0 mm corresponds to the $\lambda_{short}$ bandpass peak at the beginning of the filter. The error bars are derived from the finite aperture size and the uncertainty in the location of the aperture. \textit{Left}: Non-AR coated LVBF sample. Black line: Expected fabricated bandpass peak wavelengths. Purple dots: Bandpass peak wavelengths for the 300 K measurements. Red dots: Bandpass peak wavelengths for the 5 K measurements. \textit{Right}: AR coated LVBF sample. Black line: Expected fabricated bandpass peak wavelengths. Purple diamonds: Bandpass peak wavelegnths for the 300 K measurements. Red diamonds: Bandpass peak wavelengths for the 5 K measurements.}
\label{fig:LVF_meas_BPpeak}
\end{figure}

 We also compared the expected wavelengths of the bandpass peaks to the measured wavelengths of the bandpass peaks along the length of the LVBF. Fig. \ref{fig:LVF_meas_BPpeak} shows how the measured and expected wavelengths of the bandpass peaks vary along the filter for the non-AR coated and AR coated LVBF. The error bars take into account the 0.5 mm tolerance in the sample holder location and the 2 mm aperture. In the non-AR coated measurements there is an average difference of 0.9 $\mu$m at 300 K and 0.6 $\mu$m at 5 K between the measured and expected bandpass peak wavelengths. The AR coated measurements are more similar at 300 K and 5 K with an average difference of 1.2 $\mu$m between the measured and expected bandpass peak wavelengths. The slightly larger deviation in comparison to the non-AR coated sample could be due to fabrication variations in the cross-slot features. This should not be a problem for the BEGINS LVBF, since there is 5 \% design tolerance in the BEGINS bandpass peak wavelength. 

\par Differences in the bandpass peak wavelengths between the 5 K and 300 K measurements are also observed. There is an average difference of 0.7 $\mu$m between the 5 K and 300 K non-AR coated measured bandpass peak wavelengths and an average difference of 0.2 $\mu$m between 5 K and 300 K AR coated measured bandpass peak wavelengths. When the material is cooled there are multiple changes occurring to the LVBF, such as the resistivity of the gold decreasing, thermal contraction of the materials, and changes in the relative permittivity of the Si and AR coatings. Of these the thermal contraction of the materials and changes in the relative permittivity would effect the bandpass peak wavelength. However, thermal contractions of the materials would be negligible and not effect the bandpass peak. The Si relative permittivity changes from 11.7 to $\sim$11.6 when cooled \cite{wollack2020infrared}, which would also not significantly change the bandpass peak wavelength. We also expect the changes in the AR coating relative permittivity to be minimal. Another factor that would cause a difference in the bandpass peak wavelengths between 5 K and 300 K is the initial placement of the sample holder. If the sample holder was not returned to the same initial location for the 5 K measurements after the 300 K measurements were made then the bandpass peaks recorded would be at different locations on the LVBF. To verify that the initial placement of the sample holder is the main issue we plan to make measurements at one location on the LVBF at 300 K then at 5 K. This will eliminate moving the sample holder. 

\par
\begin{figure}[h!]
\includegraphics[width=\linewidth]{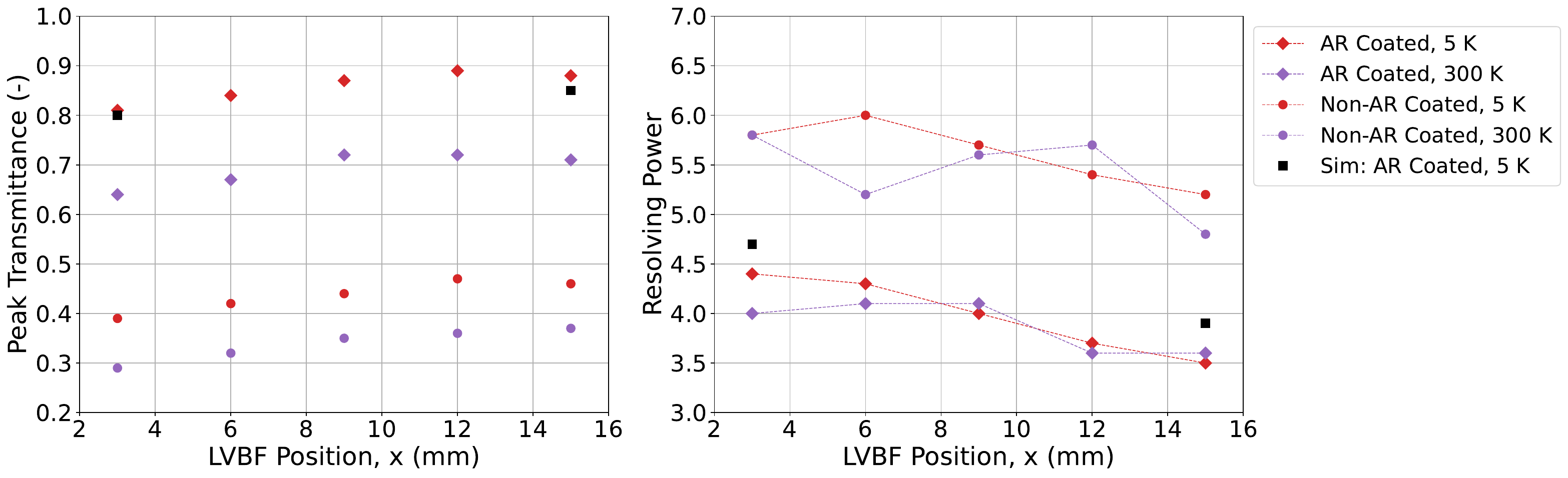}
\caption{\textit{Left}: Dots and Diamonds: FTS Measured peak transmittance as a function of the LVBF position, where x = 0 mm corresponds to the $\lambda_{short}$ bandpass peak at the beginning of the filter. Black Squares: Simulated peak transmittance for the LVBF at 5 K with AR coatings at x = 3 mm and 15 mm. Corresponding legend found on right side of Fig. \ref{fig:LVF_meas_TranR}. \textit{Right}: Dots and Diamonds: Measured resolving power as a function of the LVBF position, where x = 0 mm corresponds to the $\lambda_{short}$ bandpass peak at the beginning of the filter. Black Squares: Simulated resolving power for the LVBF at 5 K with AR coatings at x = 3 mm and 15 mm.}
\label{fig:LVF_meas_TranR}
\end{figure}

 Finally, we discuss the results of the peak transmission and resolving power of the bandpasses along the LVBF. It is important to achieve high peak transmission, such that the filters do not limit the amount of power received by the detectors. Fig. \ref{fig:LVF_meas_TranR} shows how the measured peak transmission and resolving power vary along the filter. The red and purple markers represent the peak transmission and resolving power from measurements. The resolving power was determined by, $R=\frac{\lambda_0}{\Delta \lambda_{FWHM}}$, where $\lambda_0$ is the bandpass peak wavelength and $\Delta \lambda_{FWHM}$ is the full-width at half maximum of the bandpass peak. For all measurements the peak transmission increases along the length of the filter and the resolving power decreases. Simulations of the fabricated filter's cross-slot dimensions shown in Fig. \ref{fig:LVF_fbdResults} at x = 3 mm and 15 mm were performed for an AR coated filter at 5 K (black squares in Fig. \ref{fig:LVF_meas_TranR}), to determine if the simulated resolving power and peak transmission follow the same trend as the measurements. The simulated peak transmission was in well agreement with the measured peak transmission. The simulations show that the resolving power decreases by $\sim$1 from x = 3 mm to x= 15 mm on the LVBF. The decrease in resolving power along the length of the filter is in well agreement with the decrease shown in the measured resolving power. At both 300 K and 5K the resolving power of the AR-coated sample varies from $\sim$ 4.5 to 3.5 along the length of the filter, and for the non-AR coated sample it varies from $\sim$ 6 to 5. Ideally the resolving power would stay constant for a very thin perfect conductor sheet. However, since there are resistive losses and the gold has a finite thickness the resolving power is wavelength dependent in the far-IR (THz frequencies) causing it to decrease as the bandpass peak increases in wavelength along the LVBF. Therefore, only scaling the cross-slot dimensions along the length of the filter does not ensure a constant resolving power.

\par The measured resolving power is on average $\sim$0.3 less than the simulated resolving power. One reason this occurs is because the varying bandpasses within the 2 mm FTS aperture decreases the resolving power, whereas the simulation assumes a uniform bandpass. The decrease in resolving power due to the aperture can be estimated by integrating over the product of the aperture and the bandpasses within the aperture. Performing this calculation over a 2 mm aperture estimates that the resolving power should only degrade by $\sim$0.15. Therefore, the aperture size partially contributes to the decrease in the measured resolving power. Other contributions may include the aperture not placed directly at the beam focus of the FTS, other deviations between the design and fabricated cross-slot dimensions that were not imaged, and the estimated relative permittivity of the Parylene-C AR coating used in the simulations differing from that of the Parylene-C deposited on the filters. The AR coating changes the effective capacitance of the LVBF, which affects the resolving power. We did not obtain an empirical value of the Parylene-C relative permittivity from HZO to confirm with simulations if the difference was significant enough to change the resolving power.

\section{Hydrogenated Amorphous Silicon LVBF}
\label{sec:asi_LVF}
In this section we present our results on the a-Si:H LVBF, where a 1 $\mu$m thick layer of a-Si:H was deposited on a 17 mm long LVBF for target bandpass peaks from wavelengths of 24 to 36 $\mu$m. The purpose of this investigation was to determine if the addition of the a-Si:H shifted the undesired out-of-band transmission (or higher-order side bands) exhibited by metal-mesh filters away from the bandpass peak. If the higher-order side bands of each bandpass peak are shifted far enough they can be filtered out with a low-pass filter without filtering out the bandpasses required for the LVBF. This sample was fabricated at Jet Propulsion Laboratory and made with aluminum rather than gold. We describe the fabrication method, the measurement results, compare the results to the LVBF discussed above and compare measurements to simulation.

\subsection{Fabrication}
\label{sec:aSifabrication}
The a-Si:H LVBF was fabricated on a high-resistivity DSP silicon wafer. The wafer was etched in a vapor phase hydroflouric etch tool to remove the native oxide and introduced into the load locked ultra high vacuum deposition system. The aluminum layer was direct current magnetron sputtered from a 6-inch high purity planar target to a thickness of 500 nm. The cross-slots were patterned using a deep ultraviolet (DUV) (248 nm) stepper and etched using chlorine chemistry in an Inductively Coupled Plasma (ICP) etcher. The sheet resistance of the aluminum was not measured but previous measurements from JPL indicate that sheet resistance is expected to be 3.70 $\mu \Omega$-cm at 300 K. The a-Si:H was deposited in an Inductively Coupled Plasma Enhanced Chemical Vapor Deposition (ICP-PECVD) system to a thickness of about 1 $\mu$m. For this set of filters the bandpass increased every 170 $\mu$m. Table \ref{tab:asi_cross_params} shows the design cross-slot dimensions for the bandpass peak at 24 $\mu$m and at 36 $\mu$m. The addition of the a-Si:H allowed for the cross-pitch to be decreased by a factor of 0.70 when compared to the cross-pitch of the LVBF without the a-Si:H (Table \ref{tab:cross_params}). The decrease in cross-pitch should result in higher-order side bands that are shifted further away from the bandpass peak.

\begin{table}[h!]
    \centering
    \begin{tabular}{|c|c|c|}
        \hline
         Parameter ($\mu$m) & $\lambda_0= 24$ $\mu$m & $\lambda_0= 36$ $\mu$m\\
         \hline
         \hline
         g & 4.32 & 6.48\\
         \hline
         K & 3.58 & 5.38\\
         \hline
         B & 0.80 & 1.20\\
         \hline
    \end{tabular}
    \caption{The cross-slot dimensions for bandpass peaks at 24 $\mu$m and 36 $\mu$m for a 17 mm long a-Si:H LVBF. The cross-slot dimensions for the bandpass peak at 24 $\mu$m were optimized through simulations and scaled up to 36 $\mu$m to span the full bandpass wavelength range of the a-Si:H LVBF.}
    \label{tab:asi_cross_params}
\end{table}

\begin{figure}[h!]
\includegraphics[width=\linewidth]{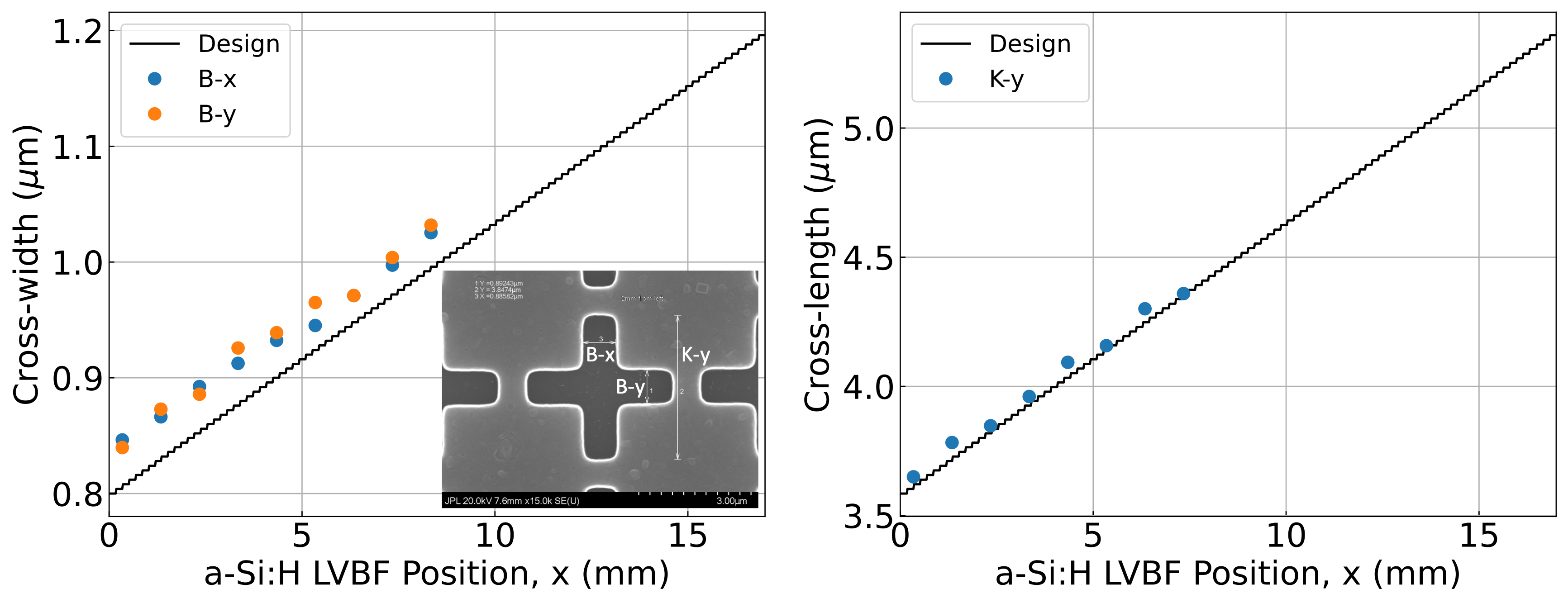}
\caption{Plots showing the results of the fabricated cross-slot dimensions as a function of the a-Si:H LVBF position, where x = 0 mm corresponds to the $\lambda_{short}$ bandpass peak at the beginning of the filter. Measurements were made along one row of the filter at x = 0.342, 1.342, 2.342, 3.342, 4.342, 5.342, 6.342, 7.342, and 8.342 mm. \textit{Left}: The measurements of the cross-width in the horizontal (B-x) and vertical (B-y) directions. \textit{Right}: Measurements of the cross-length in the vertical direction. The solid line in the plots are the design dimensions.}
\label{fig:aSiLVF_fbdParams}
\end{figure}

Measurements of the cross-slots were made on half of an a-Si:H LVBF sample after fabrication along one row of the filter at x = 0.342, 1.342, 2.342, 3.342, 4.342, 5.342, 6.342, 7.342, and 8.342 mm. The results of the measurements are shown in Fig. \ref{fig:aSiLVF_fbdParams}. The left plot shows that the cross-widths deviated from design by an average error of 4\%. The cross-lengths shown in the right plot were similar to the design dimensions with an average error of 1\%. The SEM images of the cross-slots displayed rounding in the outer and inner corners of the cross-slots with an estimated radii of a quarter of the cross-width. The deviations from the design cross-slot dimensions will result in bandpass peaks that differ from the design bandpass peaks of 24 to 36 $\mu$m.

\subsection{Filter Measurement vs Simulations: Results and Discussion}
\begin{figure}[h!]
\center
\includegraphics[width=1\linewidth]{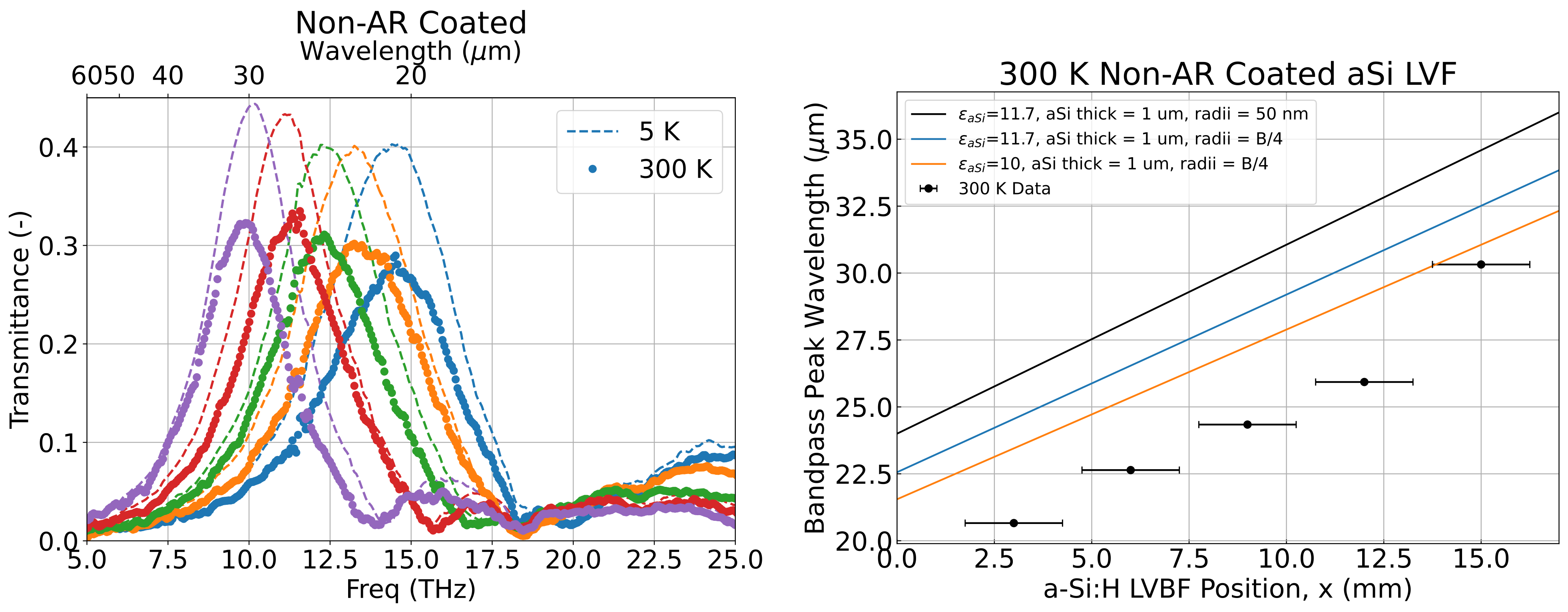}
\caption{\textit{Left}: FTS measured transmission spectra of the a-Si:H LVBF at x = 3, 6, 9, 12 and 15 mm. Measurements of the non-AR coated LVBF at 300 K (dotted lines) and 5 K (dashed lines). \textit{Right}: Plot of the wavelengths of the bandpass peaks as a function of the a-Si:H LVBF position, where x = 0 mm corresponds to the $\lambda_{short}$ bandpass peak at the beginning of the filter. The error bars are derived from the finite aperture size and the uncertainty in the location of the aperture. Black line: Expected design bandpass peak wavelengths with cross-slot corners of 50 nm radii. Blue line: Expected bandpass peak wavelengths with cross-slot corners with a radii of a quarter of the cross-width. Orange line: Expected bandpass peak wavelengths with cross-slot corners with a radii of a quarter of the cross-width and $\epsilon_{a-Si:H}=10$. Black dots: Bandpass peak wavelengths of the 300 K measurements.}
\label{fig:aSiLVF_meas}
\end{figure}

Measurements on a non-AR coated a-Si:H LVBF were performed in the same manner described in Sec. \ref{sec:MeasDesc}. The left plot in Fig. \ref{fig:aSiLVF_meas} shows the FTS measured transmission along the filter at 300 K and 5 K. The peak transmission is $\sim$30 \% at 300 K and $\sim$40-45\% at 5 K which is comparable to the non-AR coated LVBF peak transmission shown in Fig. \ref{fig:LVF_meas}. The spectral distance between the bandpass peak to the $TE_{01/10}/TM_{01/10}$ frequency cutoff is an average of 7.1 $\mu$m over all measured bandpasses. The addition of the a-Si:H increases the distance by 2.4 $\mu$m when compared to the LVBF discussed in the previous section. It also makes the bandpass peak transmission profile more symmetric, because the higher-order side bands are shifted far enough that the first higher-order side band does not interfere with the bandpass transmission on the high-frequency side. The addition of the a-Si:H was successful in shifting the higher-order side bands further from the bandpass peaks but not at far as necessary. The first higher-order side band must be shifted at least an octave away from the bandpass peak to higher frequencies, in order to successfully filter them out with a low-pass filter. 

Although the addition of the a-Si:H shows initial promising results the bandpass peak wavelength was not well predicted by the simulations when compared to the measurements. Simulations for the design cross-slot dimensions were performed under the assumption that the a-Si:H was 1 $\mu$m thick with a relative permittivity of $\epsilon_{a-Si:H}=11.7$. It was also assumed that the cross-slot corners would be rounded with a radii of 50 nm. This would have resulted in the target bandpass peak wavelengths from 24 to 36 $\mu$m as shown by the black line in the right plot of Fig. \ref{fig:aSiLVF_meas}. However, the measured bandpass peak wavelengths along the filter vary from 20.7-30.3 $\mu$m (Fig. \ref{fig:aSiLVF_meas} \textit{Right} black dots). The solid blue line shows the expected bandpass peak wavelengths if the cross-slot dimensions are modified to match the fabricated cross-slot dimensions. This includes rounding of the cross-slot corners with a radii of a quarter of the cross-width and the cross-slot dimension measurements in Fig. \ref{fig:aSiLVF_fbdParams}. Including these modifications shifted the bandpass peak to shorter wavelengths but not a significant amount to match measurements. This suggests that the adopted relative permittivity of the a-Si:H in simulation of the array geometry was differed from the material employed.

Optical measurements made at JPL demonstrated that the relative permittivity was closer to 10. The orange line shows the simulation results with $\epsilon_{a-Si:H}=10$, rounding of the cross-slot corners with a radii of a quarter of the cross-width and the cross-slot dimension measurements in Fig. \ref{fig:aSiLVF_fbdParams}. This change shifted the bandpass peak to shorter wavelengths but still not enough to match measurement with an average difference of 9\%. The difference may still be due to variations in the cross-slot along the filter that deviate from design, since only one row of cross-slot dimensions were measured along the filter. Further investigation is required of the a-Si:H and cross-slot dimensions to better predict the bandpass peak wavelength and over all transmission profile of the a-Si:H LVBF. The average resolving power for the 300 K and 5 K measurements was found to be, R$\approx$3.  If the $\epsilon_{a-Si:H}=10$, the simulations predict that R=4. The simulation also predicts that the measured peak transmission is less than the simulated peak transmission by $\sim$ 10\%. Further investigation of the dielectric properties of the a-Si:H deposited on the LVBF is still required to understand the differences between the simulations and measurements.

\section{Conclusion and Future Work}
\label{sec:Con}
We have successfully fabricated and measured a non-AR coated and AR coated LVBF and a a-Si:H LVBF as preliminary work towards defining the instruments bands of BEGINS. We demonstrated high peak transmission of $\sim$ 80-90 \% for the AR coated LVBF at cryogenic temperatures (5 K). Comparisons between simulations and measurements show how the filter's response is sensitive to changes in the
design cross-slot features when fabricated \cite{melo2012cross, page1994millimeter, perido2022cross}. Incorporating the changes in the cross-slot features after fabrication allows us to better model our measurements. For LVBFs this is more challenging since the cross-slot dimensions are changing across the filter. The addition of a thin layer of a-Si:H deposited on the metal-mesh filter did increase the spectral distance between the bandpass peak and higher-order side bands but not enough to filter out with a low-pass filter. Simulations of the a-Si:H LVBF did not match measurements well because we need a better measurement of the relative permittivity of the of a-Si:H. Our next steps are to fabricate prototype LVBFs with the resolving powers required for BEGINS (R$\geq$7.5), to further investigate the best method to decrease the effects of the high frequency higher-order side bands, and investigate the dielectric properties of a-Si:H to fabricate an a-Si:H LVBF with AR coatings. The resolving power will be increased by using UV stepper lithography which will enable better control over the cross-slot dimensions and narrower slots. The bandwidth decreases as the ratio of g/B increases, so narrower slots lead to an increase in resolving power. To suppress the higher-order side bands we will explore mesh metallization on both sides of the Si substrate, which is similar to stacking filters. Although this will decrease transmission, it will increase the resolving power. We will also explore modifying the cross-slot geometry and other metal-mesh geometries \cite{lu2018new,shahounvand2022design}.

\begin{backmatter}
\bmsection{Funding} 
Goddard Space Flight Center Internal Research and Development Award. NASA Award 80NSSC23K1598.

\bmsection{Acknowledgments} 
Joanna Perido was supported by the NASA Future Investigators in NASA Earth and Space Science Graduate Program. Nicholas F. Cothard was supported by the NASA Postdoctoral Program Fellowship at NASA GSFC.

\bmsection{Disclosures} 
The authors declare no conflicts of interest.

\end{backmatter}





\bibliographystyle{apa}
\bibliography{Optica-journal-template}



\ifthenelse{\equal{\journalref}{aop}}{%
\section*{Author Biographies}
\begingroup
\setlength\intextsep{0pt}
\begin{minipage}[t][6.3cm][t]{1.0\textwidth} 
  \begin{wrapfigure}{L}{0.25\textwidth}
    \includegraphics[width=0.25\textwidth]{john_smith.eps}
  \end{wrapfigure}
  \noindent
  {\bfseries John Smith} received his BSc (Mathematics) in 2000 from The University of Maryland. His research interests include lasers and optics.
\end{minipage}
\begin{minipage}{1.0\textwidth}
  \begin{wrapfigure}{L}{0.25\textwidth}
    \includegraphics[width=0.25\textwidth]{alice_smith.eps}
  \end{wrapfigure}
  \noindent
  {\bfseries Alice Smith} also received her BSc (Mathematics) in 2000 from The University of Maryland. Her research interests also include lasers and optics.
\end{minipage}
\endgroup
}{}

\end{document}